\documentclass{article}
\usepackage[T1]{fontenc}
\usepackage[utf8]{inputenc}
\usepackage{ismir} 
\usepackage{amsmath,cite,url}
\usepackage{graphicx}
\usepackage{xcolor}
\usepackage{booktabs}

\usepackage{tabularx}
\usepackage{multirow}
\usepackage{graphicx}
\usepackage{subcaption}
\usepackage{makecell}

\title{Beat This!\ Live -- Online Beat and Downbeat Tracking\\by Predicting the Future}

\multauthor
{Bánk Juhász$^1$ \hspace{1cm} Gerhard Widmer$^{1,2}$ \hspace{1cm} Jan Schlüter$^1$}
{$^1$ Institute of Computational Perception, Johannes Kepler University Linz, Austria\\
$^2$ LIT AI Lab, Linz Institute of Technology, Austria\\
\texttt{bank.juhasz@gmail.com, gerhard.widmer@jku.at, jan.schlueter@jku.at}}

\def\authorname{B. Juhász, J. Schlüter, and G. Widmer}

\usepackage[bookmarks=false,pdfauthor={\authorname},pdfsubject={\pdfsubject},hidelinks]{hyperref}
\usepackage[capitalise]{cleveref}

\begin{document}

\maketitle

\begin{abstract}
Beat and downbeat tracking have seen multiple improvements through recent deep learning approaches, but most target an offline setting, precluding their use in real-time scenarios.
We adapt a recent offline beat tracking model, \textit{Beat~This!} \cite{foscarin2024beatthisaccuratebeat}, for online, low-latency operation and propose \textit{Beat~This!~Live}, a causal variant designed for real-time beat and downbeat tracking.
We present the architectural modifications required to obtain a causal system suitable for streaming inference and
measure their impact on accuracy.
By training our model with temporally shifted labels, it produces predictions slightly ahead of the current frame,
which not only reduces latency, but enables much more effective peak-picking.
Evaluation on the GTZAN \cite{marchand2015swing} dataset shows the F1-score improves from 80.6 to 86.7 for beats and 56.5 to 74.1 for downbeats compared to \textit{BeatNet+} \cite{heydari2024beatnet+}, the current state of the art in online beat tracking.
\end{abstract}

\section{Introduction}\label{sec:introduction}

Beat and downbeat tracking constitute fundamental problems in MIR, aiming to estimate the temporal positions of beats and downbeats in a music recording. The beat corresponds to the regular pulse underlying the music, while the downbeat marks the first beat in a musical bar and indicates the metrical structure. However, many of the most accurate beat tracking systems operate in an offline setting, where the
entire audio signal is available before processing begins. These models often rely on large temporal context or bidirectional sequence modeling, which allows them to incorporate information from both past and future frames. While effective for offline analysis, such approaches are unsuitable for real-time scenarios where beats and downbeats must be predicted continuously for a live audio stream and future context is unavailable.

In this work, we take a recent offline model as a basis to develop a competitive online model (but not yet a real-time implementation).
For the basis, we pick Beat This!\ \cite{foscarin2024beatthisaccuratebeat}, as it provides near state-of-the-art performance and is fully open source. In pursuit of an effective online version, we remove any source of future information that could leak into the model’s predictions, resulting in a \emph{causal} model.

However, as a prerequisite to operate in \textit{real-time}, causality alone is not enough. The model must also be able to deliver predictions with sufficiently low latency. As humans can generally detect an audio latency of \textasciitilde{}50 ms (with some being far more sensitive than that) \cite{schmid2024measuring}, minimizing it in any real-time audio prediction task becomes crucial. 

We address this challenge by exploiting the temporal regularity of music and adapt Meier et~al.'s \cite{meier2024real} idea of a \textit{beat lookahead} by shifting the labels back by a few frames during training. Similarly to how humans tap along to a metronome by \textit{anticipating} beats \cite{patel2014evolutionary}, we predict beats and downbeats several frames before they occur, leaving ample time for inference computations (\cref{fig:shift}). Not only does this enable zero- or even negative-latency predictions, but this also makes postprocessing decisions more effective by allowing the model to peek ahead of the current frame.

Summarizing, this work makes three primary contributions. First, it demonstrates a systematic approach for transforming the Beat This!\ architecture into a causal variant suitable for online beat and downbeat tracking. Second, it provides an empirical analysis of how causal architectural modifications and postprocessing strategies affect tracking accuracy while keeping latency low enough to facilitate real-time applications. Third, it introduces the Shifted Causal Local Max postprocessing method, which enables effective peak detection in causal systems through label shifting and predictive inference.

\section{Related Work}\label{sec:related_work}

Although there have been real-time beat tracking efforts already in the 1980s \cite{dannenberg1987following}, the first real-time methods that processed real audio signals instead of synthetic MIDI performances were developed by Goto and Muraoka in 1994 \cite{goto1994beat}. Their system extracted onset information across multiple frequency bands and maintained several concurrent hypotheses through a set of agents, each predicting future beat positions according to different strategies. This multi-agent approach would be refined by several works in the years to come \cite{goto1994beat,goto1995real,goto1996beat,oliveira2010ibt}. In parallel with this line of work, Scheirer \cite{scheirer1998tempo} introduced a contrasting signal-processing approach based on \textit{psychoacoustics}. He used bandpass filters and banks of parallel comb filters to predict beats causally. Klapuri et~al.'s model \cite{klapuri2005analysis}, built on Scheirer's work, jointly infers tatum, tactus (beat), and measure (downbeat).

More recently, deep learning approaches have become popular and have seen multiple improvements in the last decade. Gkiokas \& Katsouros \cite{gkiokas2017convolutional} were the first to use CNNs for beat tracking, successfully tackling the online task right away. Heydari et~al. \cite{heydari2021beatnet} proposed BeatNet, a CRNN with particle filtering for simultaneously predicting beats and downbeats. BEAST \cite{chang2024beast} was able to outperform BeatNet with a streaming transformer architecture which processes the input in blocks with some context inheritance mechanisms to retain information between blocks. The current state of the art, however, is a recent extension of BeatNet: the authors built on their original model to create BeatNet+ \cite{heydari2024beatnet+}, where the LSTM block now consists of four layers instead of two and they use an additional branch during training with slightly altered input to increase robustness. In a return to more classical signal processing approaches, Meier et~al. \cite{meier2024real} achieves competitive performance while maintaining zero latency for predicting beats.

The offline task generally receives more attention and there has been no shortage of improvements in this realm in the past few years. Hung et~al.'s \cite{hung2022modeling} approach combined a transformer termed SpecTNT with a temporal convolution network, and used the Dynamic Bayesian Network (DBN) of Böck et~al.\ \cite{bock2016dbn} as a postprocessor to achieve state-of-the-art results. Currently, a major research direction involves the adaptation of pretrained music foundation models (e.g., MERT\cite{yizhimert} and MusicFM\cite{won2024foundation}) through parameter-efficient fine-tuning methods like HingeNet \cite{ru2025hingenet}, which incorporates a harmonic-aware mechanism to capture harmonic shifts, or BeatFM \cite{ru2025beatfm}, which uses a multi-dimensional semantic aggregation module to focus on temporal, frequency, and channel domains. Moreover, recent self-supervised learning frameworks, such as Knowledge-Driven Multiple Hypothesis Learning (KD-MHL) \cite{gagnere2025controlling}, have been introduced to handle rhythmic ambiguity and data scarcity by considering multiple plausible interpretations during pre-training. We choose Beat This!\ \cite{foscarin2024beatthisaccuratebeat} as our foundation, as it is close to the state of the art, and completely open source.

\section{Methodology}\label{sec:methodology}

 Our implementation and architecture is directly based on the original Beat This!,\ which processes precomputed $T \times 128$ spectrograms using a rotary transformer with two task heads: one for beats and one for downbeats, producing $T \times 2$ probability estimates. For peak picking, it uses a simple method which finds local probability maxima to emit beat and downbeat signals \cite{foscarin2024beatthisaccuratebeat}. In the following, we present all modifications to the original model to make it causal and able to operate with minimal or no latency. To aid experimentation and to make the model more lightweight, the vast majority of the presented model variants use the smaller transformer dimension (128) found in the paper \cite{foscarin2024beatthisaccuratebeat}.

\subsection{Causality}\label{subsec:causality}

The necessary architectural changes affect all layers which operate over the time dimension: all convolutional layers, the time-directed partial transformers, and the transformer blocks. Beat This!\ uses 2D convolutions in its \textit{stem} and \textit{frontend blocks}, which convolve over frequency and time \cite{foscarin2024beatthisaccuratebeat}. We replace these layers with causal convolutions instead \cite{van2016wavenet}, using padding and cropping to ensure causality. Beat This!\ also uses batch normalization, which is not causal during training, but is replaced with a fixed framewise linear transformation for inference that is causal.

The frontend blocks use a combination of partial transformers: the first one is frequency-directed, treating each time frame as a sequence of length $F$ -- the number of frequency bands \cite{foscarin2024beatthisaccuratebeat}. As attention over frequency bands does not violate causality, these were left unchanged. However, the second one operates over time and treats each frequency band as a sequence of length $T$ -- the number of frames \cite{foscarin2024beatthisaccuratebeat}. Here, as well as in all six transformer blocks, we replace bidirectional attention with causal self-attention instead \cite{vaswani2017attention}.

\subsection{Streaming}\label{subsec:method:streaming}

For inference, Beat This!\ processes the input with full attention over non-overlapping 30 s blocks \cite{foscarin2024beatthisaccuratebeat}. While this setup works well for the offline task and keeps the computational costs low, for the online task, it would lead to poor results, as at the beginning of every new block, the model would not have any past context for its prediction. To avoid these resets,
we could use causal full self-attention, but this would result in a growing computational complexity for every additional input frame (complexity would be linear in the number of accumulated past frames) until eventually processing could not keep up with the input frame rate.

Instead, we adopt causal sliding-window attention \cite{vaswani2017attention,beltagy2020longformer}, where only a fixed local window of preceding tokens attends to each position rather than the entire history. This way the computational complexity is constant across incoming input frames, controlled by the window size $W$, while still capturing short-range dependencies.

\begin{figure}[t]
    \centering

    \begin{subfigure}{\linewidth}
        \centering
        \begin{minipage}{0.28\linewidth}
            \captionsetup{justification=raggedright, singlelinecheck=false}
            \caption{Beats}
        \end{minipage}
        \hfill
        \begin{minipage}{0.68\linewidth}
            \includegraphics[width=\linewidth]{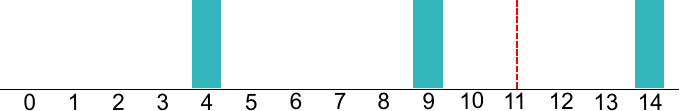}
        \end{minipage}
    \end{subfigure}

    \vspace{0.3em}

    \begin{subfigure}{\linewidth}
        \centering
        \begin{minipage}{0.28\linewidth}
            \captionsetup{justification=raggedright, singlelinecheck=false}
            \caption{Unshifted Causal}
        \end{minipage}
        \hfill
        \begin{minipage}{0.68\linewidth}
            \includegraphics[width=\linewidth]{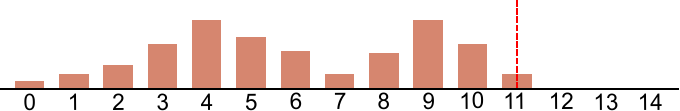}
        \end{minipage}
    \end{subfigure}

    \vspace{0.3em}

    \begin{subfigure}{\linewidth}
        \centering
        \begin{minipage}{0.28\linewidth}
            \captionsetup{justification=raggedright, singlelinecheck=false}
            \caption{Shifted Causal (3 frames)}
        \end{minipage}
        \hfill
        \begin{minipage}{0.68\linewidth}
            \includegraphics[width=\linewidth]{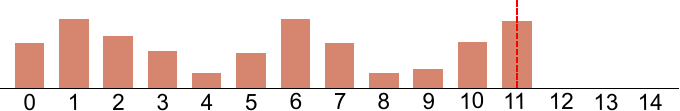}
        \end{minipage}
    \end{subfigure}

    \caption{
        Example estimates for the unshifted (b) and shifted (c) causal models with the corresponding true label (a). At frame 11 (red line), the shifted model (c) produces its estimate for frame 14 (3-frame lookahead). This provides additional time for computational latency as the model must only emit this prediction at frame 14 -- potentially allowing for zero- or negative-latency predictions.
    }
    \label{fig:shift}
\end{figure}

\subsection{Latency \& Postprocessing}\label{subsec:latenc_n_postprocessing}

Although the model operates in a causal setting, the input representation is derived from a centered STFT with a 1024-sample window, resulting in an implicit lookahead of approximately 23 ms at 22.05 kHz. This introduces a non-negligible feature-level latency, which is not accounted for in the model architecture but is compensated in postprocessing. This is a consequence of relying on the spectrograms published by the authors of Beat This!\ without alterations; not all training datasets have publicly available audio data to recalculate the spectrograms causally. 

To combat this inherent latency, and to allow time for computing predictions, we train models with shifted labels: true beats and downbeats are shifted back by a set number of frames during training (\cref{fig:shift}). This forces the model to \emph{anticipate} beats and downbeats before they occur. As this is a somewhat harder task, some performance drop is expected, but there is also an upside: we gain time to make better postprocessing decisions.

\begin{figure}[t]
    \centering

    \begin{subfigure}{\linewidth}
        \centering
        \begin{minipage}{0.28\linewidth}
            \captionsetup{justification=raggedright, singlelinecheck=false}
            \caption{Offline}
            \label{subfig:postp_a}
        \end{minipage}
        \hfill
        \begin{minipage}{0.68\linewidth}
            \includegraphics[width=\linewidth]{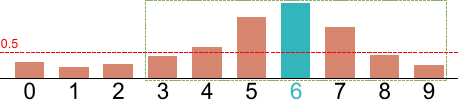}
        \end{minipage}
    \end{subfigure}

    \vspace{0.3em}

    \begin{subfigure}{\linewidth}
        \centering
        \begin{minipage}{0.28\linewidth}
            \captionsetup{justification=raggedright, singlelinecheck=false}
            \caption{Causal Thresholding at 0.5}
            \label{subfig:postp_b}
        \end{minipage}
        \hfill
        \begin{minipage}{0.68\linewidth}
            \includegraphics[width=\linewidth]{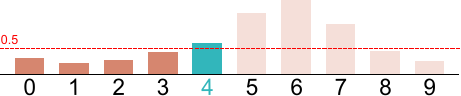}
        \end{minipage}
    \end{subfigure}

    \vspace{0.3em}

    \begin{subfigure}{\linewidth}
        \centering
        \begin{minipage}{0.28\linewidth}
            \captionsetup{justification=raggedright, singlelinecheck=false}
            \caption{Causal Thresholding at 0.8}
            \label{subfig:postp_c}
        \end{minipage}
        \hfill
        \begin{minipage}{0.68\linewidth}
            \includegraphics[width=\linewidth]{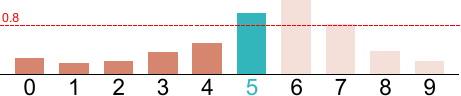}
        \end{minipage}
    \end{subfigure}

    \vspace{0.3em}

    \begin{subfigure}{\linewidth}
        \centering
        \begin{minipage}{0.28\linewidth}
            \captionsetup{justification=raggedright, singlelinecheck=false}
            \caption{Shifted Causal Local Max}
            \label{subfig:postp_d}
        \end{minipage}
        \hfill
        \begin{minipage}{0.68\linewidth}
            \includegraphics[width=\linewidth]{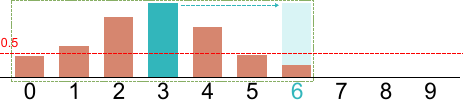}
        \end{minipage}
    \end{subfigure}

    \caption{
    Comparison of postprocessing strategies.
    Offline Beat This!\ picks local maxima above a threshold (a), which requires access to future predictions.
    Causal thresholding reports the first activation above a threshold (b, c). A low threshold reports events too early (b). Raising the threshold reports events later (c), but may miss events completely.
    Training with shifted labels moves all predictions to occur early (d), allowing to pick local maxima without breaking causality.
    }
    \label{fig:postp}
\end{figure}

The postprocessing method used in Beat This!\ picks local 7-frame probability maxima which are larger than the threshold of 0.5 (\cref{subfig:postp_a}). Additionally, downbeat predictions are snapped to the closest beat prediction which results in more musically meaningful predictions \cite{foscarin2024beatthisaccuratebeat}. While proven to be effective for the offline task, this scheme clearly violates causality, as detecting local maxima relies on peeking ahead of the current frame. Snapping downbeats to beats is similarly problematic.

For models which were trained without shifting labels (i.e., which predict the current frame), we use simple causal thresholding with a cooldown (\cref{subfig:postp_b,subfig:postp_c}). Iterating over frames, if the beat probability estimate is above a predefined threshold, we emit a beat. In case a beat is found, we check the probability estimate for a downbeat as well. As predicting beats is generally the easier task of the two, and it being a superset of downbeats, it makes sense to gate downbeat emissions by beats. The addition of a cooldown period after an emission avoids rapid misfires.

Naturally, both the probability threshold and the number of cooldown frames are crucial, as we can balance precision and recall by tuning them. Too low of a threshold causes the model to fire too soon and the real beat frame may fall in the cooldown period, while choosing a value too high will result in missing beats entirely. A well-selected threshold can nudge the emissions closer to the top of the wide, bell-shaped peaks that tend to form around true labels, thereby producing more accurate predictions (\cref{subfig:postp_b,subfig:postp_c}).

In the case of models where we produce predictions for the current frame, the decision about whether to emit a beat or a downbeat signal must be made immediately, otherwise we incur latency. This is fundamentally different for label-shifted models, as we are no longer forced to emit signals at the moment of prediction: we may choose to wait and take the predictions for upcoming frames into account. Drawing inspiration from Meier et~al.'s \cite{meier2024real} beat lookahead, we propose \textit{Shifted Causal Local Max} (SCLM) to exploit this small peek into the future while preserving causality (\cref{subfig:postp_d}).

For label-shifted models with shift $s$, instead of making an immediate decision at frame $t$, we consider a symmetric window $w$ of size $2s+1$, centered at $t$, i.e., frames $[t - s, \ldots, t + s]$. A beat is emitted at time $t$ if the central prediction exceeds a threshold and is the maximum within this window. This allows the model to defer decisions until sufficient future evidence is available, effectively approximating the minimal peak picking of the offline model.

Shifted-label models also allow for a more flexible downbeat postprocessing strategy. Using SCLM, we can elegantly gate downbeat predictions by positive beat signals and improve their prediction accuracy at the same time. If we identify a beat frame and there is a downbeat logit \textit{anywhere} in the window above the threshold, we emit a downbeat as well. This is conceptually the same as the downbeat-snapping of the offline postprocessing method, without sacrificing causality or latency.

By choosing a window s.t. $w < 2s+1$, we can sacrifice some of the \textit{foresight} to further reduce latency to zero, or predict with negative latency.

\section{Experiments}\label{sec:experiments}

We perform experiments on each of the aspects discussed above: causality, streaming, and postprocessing. Finally, we compare results to the state of the art.

\subsection{Training \& Evaluation}\label{subsec:trainig_n_datasets}

The datasets used for training, validation, and testing are identical to those used in the original Beat This!\ \cite{foscarin2024beatthisaccuratebeat}, and we exactly follow its training and evaluation setup, using the published code and data.
The training data encompasses 4556 tracks from 15 datasets.
A 15\% split of the training was used as a validation set, taking care not to have different versions of the same track appearing in different splits, thus preventing information leakage.
The GTZAN \cite{marchand2015swing} dataset's 993 pieces are used as a held-out test set, apart from 7 examples discarded due to missing annotations.
All augmented spectrograms and data splits were provided by the authors of Beat This!, and are used without alterations.
Training is done on random 30\,s excerpts using the original hyperparameters.
For more details, we refer to the original paper \cite{foscarin2024beatthisaccuratebeat}.

Following the practice found in the literature, we report F1-scores with $\pm70$ ms of tolerance, ignoring the first 5 seconds of each piece (both standard in mir\_eval \cite{raffel2014mireval}). In addition to theoretical latency, we report Real-Time Factor (RTF), which is the ratio of total processing time and total audio duration. All RTF metrics presented in this article (including those of previous work) were calculated on GTZAN \cite{marchand2015swing}. To compute the processing time, we used a single thread of an AMD Ryzen 5 7600 CPU at 3.80 GHz clock frequency, and included the spectrogram calculation and full model inference with postprocessing. In addition, we report two continuity metrics: CML$_{t}$ and AML$_{t}$ on the validation set. All metrics presented were calculated as the mean score based on three runs using different seeds.

\subsection{Causality}\label{subsec:exp_causality}

\begin{figure}[t]
    \centering

    \begin{subfigure}{0.485\linewidth}
        \centering
        \includegraphics[width=\linewidth]{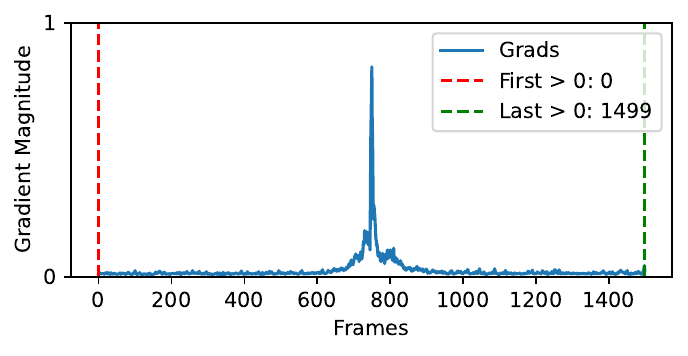}
        \caption{Beat This!\ (small)}
        \label{subfig:gradients_a}
    \end{subfigure}
    \hspace{0.001\linewidth}
    \begin{subfigure}{0.485\linewidth}
        \centering
        \includegraphics[width=\linewidth]{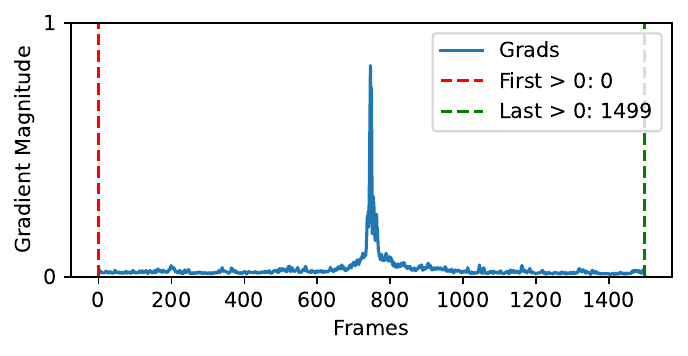}
        \caption{Causal Convolutions}
        \label{subfig:gradients_b}
    \end{subfigure}

    \vspace{0.1em}

    \begin{subfigure}{0.485\linewidth}
        \centering
        \includegraphics[width=\linewidth]{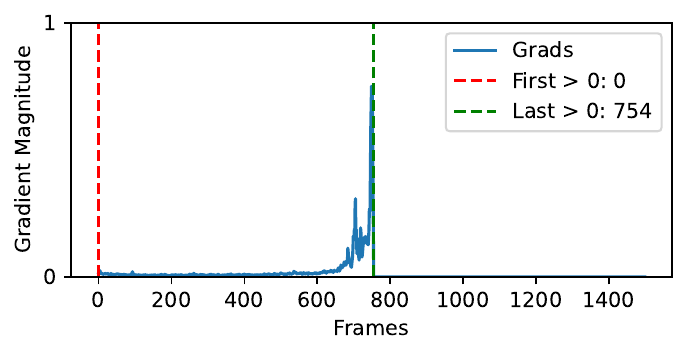}
        \caption{Causal Attention}
        \label{subfig:gradients_c}
    \end{subfigure}
    \hspace{0.001\linewidth}
    \begin{subfigure}{0.485\linewidth}
        \centering
        \includegraphics[width=\linewidth]{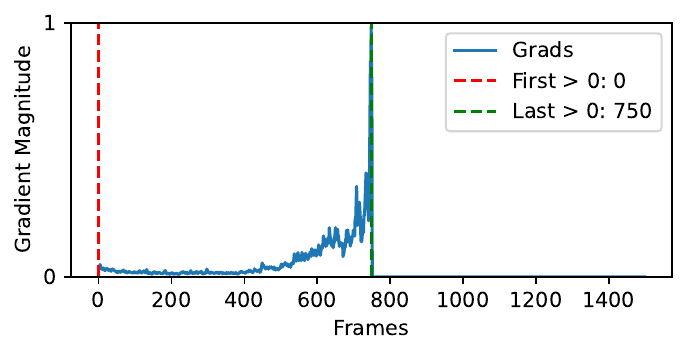}
        \caption{Causal Conv. \& Attn.}
        \label{subfig:gradients_d}
    \end{subfigure}

    \caption{Comparison of gradient responses across models for frame 750 w.r.t. an input of length 1500.}
    \label{fig:gradients}
\end{figure}

Our first changes to the original architecture are the replacement of convolutions and self-attention with causal variants.
First, we verify that with our changes, the model indeed does not include future information in its prediction for a particular spectrogram frame.
To do so, we compute the gradient of the model's prediction for the central frame of its 1500-frame input with respect to all input frames, for a random example. \cref{subfig:gradients_a} shows the summed gradients of the offline model. Clearly, it relies on context from the full input excerpt. \cref{subfig:gradients_b} shows a similar image -- in this case, we have turned convolutions causal, but the non-causal attention mechanism dominates. In \cref{subfig:gradients_c}, we turn attention causal and leave convolutions in their original, offline form. As a result of this, the last frame for which the summed gradients are non-zero is frame 754 -- the model peeks 4 frames into the future as a consequence of the combined receptive field of \textit{all} convolutional layers. Finally, \cref{subfig:gradients_d} confirms causality in the case where both attention and convolutional layers are causal.

\begin{table}[t]
 \centering
 \setlength{\tabcolsep}{4pt}
    \begin{tabularx}{\linewidth}{lcc}
    \toprule
    \textbf{Model}                 & \textbf{Beat F1} & \textbf{Downbeat F1} \\
    \midrule
    Beat This!\ small            & 88.8             & 77.2 \\
    - with Causal Convolutions     & 88.9             & 76.9 \\
    - with Causal Attention        & 88.1             & 73.6 \\
    - with Causal Conv. \& Attn.        & 87.9             & 73.4 \\
    \bottomrule
    \end{tabularx}
\caption{Performance impact of causality on GTZAN. Metrics were calculated using offline postprocessing.}
\label{tab:impact_causality}
\end{table}

To analyze the impact of causality on prediction performance, we report F1-scores for the same four variants in \cref{tab:impact_causality}. We report results on GTZAN \cite{marchand2015swing}, as these will not be used for model selection. Note that these metrics are calculated using the original offline postprocessing method of Beat This!\ to ensure that the performance changes are strictly the result of reduced context, rather than a causal postprocessing method. From the results, we see that causal attention seems to hurt estimation quality the most, and this effect is more pronounced for downbeats, which are generally harder to predict.

As we expect that our modifications to address latency and causality in postprocessing will degrade predictions further, the quality achieved by the fully causal variant with offline postprocessing can be considered a practical ceiling for performance metrics: one which later variants are unlikely to achieve.

\subsection{Streaming}\label{subsec:exp:streaming}

The gradients of the causal model (\cref{subfig:gradients_d}) suggest that the immediate history is significantly more important and the influence of earlier frames in the sequence quickly starts to drop off. This would mean that we should be able to preserve much of the performance of full causal attention models even with a fixed-sized sliding-window attention variant. Nevertheless, regardless of our choice of causal variant, the lack of temporal context at the beginning of a sequence will have an adverse effect on performance.

\begin{table}
    \centering
    \setlength{\tabcolsep}{4pt}
    \begin{tabularx}{\linewidth}{l ccc ccc}
    \toprule
    & \multicolumn{3}{c}{\textbf{Beat F1}} 
    & \multicolumn{3}{c}{\textbf{Downbeat F1}} \\
    \cmidrule(lr){2-4} \cmidrule(lr){5-7}
    \textbf{Model}
    & \textbf{S1} & \textbf{S2} & \textbf{S3}
    & \textbf{S1} & \textbf{S2} & \textbf{S3} \\
    \midrule
    
    Beat This!\ small       & 91.8 & 92.8 & 92.9 & 85.7 & 86.0 & 86.0 \\
    Full Causal Attn. & 87.6 & 89.5 & 89.9 & 77.2 & 79.0 & 79.6 \\
    SW-512            & 82.5 & 84.2 & 83.6 & 74.8 & 77.9 & 77.4 \\
    SW-256            & 82.7 & 83.9 & 83.5 & 75.4 & 77.2 & 77.1 \\
    \bottomrule
    \end{tabularx}
    \caption{Segmented metrics for different model variants: S1, S2, S3 denote evaluation on the first, second, third 10-second segment of the central 30 seconds of each validation piece. SW-512 indicates sliding window attention over 512 frames. Causal models use causal thresholding as postprocessing.}
    \label{tab:segmented_metrics}
\end{table}

\begin{table*}[t]
    \centering
    \setlength{\tabcolsep}{4pt}
    \begin{tabularx}{0.975\linewidth}{l c c c c ccc ccc c}
        \toprule
        & \textbf{\multirow{2}{*}{\parbox[t]{10mm}{\centering Transf. dim.}}}
        & \textbf{\multirow{2}{*}{\parbox[t]{10mm}{\centering Win. size}}}
        & \textbf{\multirow{2}{*}{\parbox[t]{8mm}{\centering Label shift}}}
        & \textbf{\multirow{2}{*}{\parbox[t]{8mm}{\centering Look\-ahead}}}
        & \multicolumn{3}{c}{\textbf{Beat}} 
        & \multicolumn{3}{c}{\textbf{Downbeat}}
        & \textbf{{\multirow{2}{*}{\parbox[t]{12mm}{\centering Latency (ms)}}}} \\
        \cmidrule(lr){6-8} \cmidrule(lr){9-11}
        \textbf{Model}
        & \textbf{}
        & \textbf{}
        & \textbf{}
        & \textbf{}
        & \textbf{F1} & \textbf{CML$_{t}$} & \textbf{AML$_{t}$}
        & \textbf{F1}  & \textbf{CML$_{t}$} & \textbf{AML$_{t}$}
        & \textbf{}
        \\
        \midrule
        Beat This!\ small (offline) & 128 & -- & 0  & 0  & 92.0 & 83.8 & 87.7 & 84.4 & 71.4 & 78.9 & -- \\
        \cmidrule(lr){1-1}
        128-SW256     & 128 & 256 & 0  & 0 & 83.1 & 69.5 & 72.5 & 75.8 & 63.8 & 68.9 & 23  \\
        128-SW256-1-1 & 128 & 256 & 1  & 1 & 84.5 & 70.4 & 73.2 & 76.5 & 64.7 & 70.7 & 23  \\
        128-SW256-3-3 & 128 & 256 & 3  & 3 & 83.9 & 70.7 & 73.6 & 76.0 & 64.9 & 70.7 & 23  \\
        128-SW256-5-5 & 128 & 256 & 5  & 5 & 82.4 & 70.9 & 73.5 & 73.8 & 62.4 & 67.5 & 23  \\
        \cmidrule(lr){1-1}
        128-SW256-3-1$^\ast$ & 128 & 256 & 3  & 1 & 83.2  & 69.0 & 71.9 & 75.6 &	64.9 & 70.5 & \textbf{-17} \\
        128-SW256-5-3 & 128 & 256 & 5  & 3 & 81.9  & 69.8 & 72.4 & 73.6 &	62.3 & 67.3 & \textbf{-17} \\
        \cmidrule(lr){1-1}
        128-SW512-1-1$^\dagger$ & 128 & 512 & 1  & 1 & 84.7 & 71.2 & 73.8 & 76.7 &	65.2 & 71.3 & 23  \\
        \cmidrule(lr){1-1}
        512-SW512-1-1$^\ddagger$ & 512 & 512 & 1  & 1 & \textbf{86.4} & \textbf{73.9} & \textbf{76.8} & \textbf{79.2} & \textbf{68.1} & \textbf{73.7} & 23  \\
        \bottomrule
    \end{tabularx}
    \caption{Comparing different model variants on the validation set. We first test different settings for label shifting using all of the shifted frames for postprocessing lookahead (23\,ms of delay), then reserving two shifted frames to improve delay (-17\,ms of delay), then increase sliding window size, then increase transformer dimensionality. \\
    $^\ast$ Beat This! Live Zero, 
    $^\dagger$ Beat This! Live, 
    $^\ddagger$ Beat This! Live Large.}
    \label{tab:model_variants}
\end{table*}

To assess the severity of this warm-up problem in our causal models, the postprocessed predictions for the central 30-second excerpt of each validation piece are subdivided into three 10-second segments for which performance metrics are calculated separately.\footnote{Here we do not ignore the first 5 seconds, because we specifically want to evaluate the case of incomplete past context, and the context window of the 256-frame variant would be almost full after 5 seconds.}

\autoref{tab:segmented_metrics} shows the results. As expected, the offline model performs strongly already in the first segment, although the first segment is still markedly weaker for both beats and downbeats. For all causal variants, performance is indeed much worse in the first segment, for both targets, clearly indicating the impact of limited initial context. The full causal attention variant keeps improving in each segment with more context becoming available.
The sliding-window models perform slightly worse than the full attention variant, and prediction quality mostly levels out for the second and third segment, as the amount of available context is the same.
As the results for a sliding window of 256 frames are largely comparable to those of the 512-frame variant, we will move forward with the smaller window.

\subsection{Latency \& Postprocessing}

In the last set of experiments, we compare different options for shifting the training labels to improve postprocessing, reduce delay, or both.

Starting from the offline Beat This!\ model, still with a transformer dimensionality of 128, we compare it to causal sliding window attention with varying amounts of label shift (\cref{tab:model_variants}, rows 2--5). In each case, we use all of the shift as a lookahead for peak-picking in postprocessing with SCLM, resulting in each model having a delay of 23\,ms from the spectrogram (see \cref{subsec:latenc_n_postprocessing}).

We find that using even a single frame of label shift provides a substantial benefit, as opposed to the unshifted model which can only use causal thresholding without any lookahead.
However, larger shifts do not always provide an advantage over smaller ones -- the advantage starts to drop off with the model that predicts 5-frames into the future (\texttt{128-SW256-5-5}) performing worse than the unshifted one (\texttt{128-SW256}). Seemingly, there is a trade-off between making the task harder (anticipating beats and downbeats), and making better postprocessing decisions. Best F-scores are achieved with a label-shift (and lookahead) of 1 frame (\texttt{128-SW256-1-1}), suggesting that during postprocessing, this frame is enough to identify when the framewise probability estimates start to decline and additional future frames do not provide enough benefit to counteract having to anticipate beats and downbeats.
Continuity metrics are optimal for a label shift (and lookahead) of 3 frames.

To reduce latency, instead of using all of the label shift for postprocessing, we evaluate shifting by 2 more frames than used as lookahead (\cref{tab:model_variants}, rows 6--7). This reduces latency by two frames of 20\,ms, resulting in a \emph{negative} latency of -17\,ms.
A single frame of lookahead proves optimal for F-score and downbeat continuity, and three frames are optimal for beat continuity.
We pick the former model, \texttt{128-SW256-3-1}, as the best lightweight, negative latency model, and will refer to it as \textit{Beat This!\ Live Zero}.

\begin{table}[!b]
    \centering
    \setlength{\tabcolsep}{4pt}
    \begin{tabularx}{\linewidth}{lcclr}
        \toprule
        \textbf{Model} 
        & \makecell{\textbf{Beat} \\ \textbf{F1}} 
        & \makecell{\textbf{Downbeat} \\ \textbf{F1}} 
        & \makecell{\textbf{RTF}}
        & \makecell{\textbf{Lat.} \\ \textbf{(ms)}} \\
        \midrule
        Beat This!\ Live Large               & \textbf{86.7} & \textbf{74.1} & 0.08           & 23 \\
        Beat This!\ Live                     & 86.1          & 72.8          & \textbf{0.05}  & 23 \\
        Beat This!\ Live Zero                & 85.9          & 72.2          & \textbf{0.05}  & \textbf{-17} \\
        BeatNet+ \cite{heydari2024beatnet+}  & 80.6          & 56.5          & 0.06           & 20 \\
        BEAST \cite{chang2024beast}          & 80.0          & 46.8          & 0.41$^\ast$    & 46 \\
        BeatNet \cite{heydari2021beatnet}    & 75.4          & 46.5          & 0.06$^\ast$    & 20 \\
        RNN-PLP-On \cite{meier2024real}      & 74.7          & -             & -              & 12 \\
        RNN-PLP-On-Zero \cite{meier2024real} & 74.7          & -             & -              & 0 \\
        \bottomrule
    \end{tabularx}
    \caption{Comparing Beat This!\ Live variants with recent works on online beat and downbeat tracking on the GTZAN dataset. RTFs marked $^\ast$ are from \cite{chang2024beast,heydari2024beatnet+}, measured on different hardware.}
    \label{tab:comp_with_sota}
\end{table}

\begin{table*}[!t]
    \centering
    \setlength{\tabcolsep}{4pt}
    \begin{tabularx}{0.94\linewidth}{ll ccccc ccccc}
    \toprule
    Task & Model & Blues & Classical & Country & Disco & Hiphop & Jazz & Metal & Pop & Reggae & Rock \\
    \midrule
    Beats & Beat This! Live & \textbf{78.8} & \textbf{56.3} & \textbf{91.2} & \textbf{96.4} & \textbf{95.7} & \textbf{81.6} & \textbf{87.4} & \textbf{94.2} & 88.7 & \textbf{90.4} \\
    Beats & BeatNet+ \cite{heydari2024beatnet+} & 71.3 & 53.3 & 84.7 & 95.3 & 90.8 & 69.9 & 82.4 & 87.5 & \textbf{91.8} & 86.7 \\
    Downbeats & Beat This! Live & \textbf{59.5} & \textbf{40.2} & \textbf{87.0} & \textbf{90.5} & \textbf{85.6} & \textbf{63.4} & \textbf{75.3} & \textbf{89.0} & \textbf{59.9} & \textbf{76.8} \\
    Downbeats & BeatNet+ \cite{heydari2024beatnet+} & 43.7 & 19.2 & 71.4 & 80.9 & 73.9 & 38.5 & 61.0 & 70.9 & 36.6 & 65.2 \\
    \bottomrule
    \end{tabularx}
    \caption{Comparison of Beat This! Live with BeatNet+ using F1 scores for both tasks across all genres found in GTZAN.}
    \label{tab:comp_over_genres}
\end{table*}

Reconsidering the choice of the sliding window size, we train the best model of 23\,ms latency with a sliding window of 512 frames (\texttt{128-SW512-1-1}), and find that we get better results for every metric. This is our model of choice for low latency operation, \textit{Beat This!\ Live}.

Finally, we train a larger variant with the transformer dimension increased to 512 (\texttt{512-SW512-1-1}) -- the original dimensionality found in the offline Beat This!. The increased model size provides a clear benefit over smaller variants. We call this model \textit{Beat This!\ Live Large}.

\subsection{Comparison to State of the Art}\label{subsec:main_results}

\cref{tab:comp_with_sota} displays our three chosen models compared to recent online beat and downbeat trackers. All of our model variants achieve state-of-the-art results, with a significant leap forward for online downbeat tracking. Even while operating with negative latency, we improve over the previous state of the art.

To estimate real-time capabilities, we compute the real-time factor (RTF) for each of our models as well as for BeatNet+, using the public codebase and weights. RTFs for BEAST and BeatNet are taken from publications \cite{chang2024beast} and \cite{heydari2024beatnet+}, respectively, which differ in hardware and evaluation settings. Our models achieve sufficiently low RTF values under our setup, comparable to BeatNet+, suggesting that real-time deployment is feasible.

Using the BeatNet+ codebase and weights, we can also perform a per-genre performance comparison (\cref{tab:comp_over_genres}). With one exception, our method outperforms the current state of the art across all genres for both tasks, with particularly large improvements in downbeat tracking in the most challenging categories: classical, jazz, and reggae.

\section{Conclusion \& Outlook}\label{sec:conclusion}

This article proposes a novel online beat and downbeat tracking system, Beat This!\ Live, derived from the state-of-the-art offline model Beat This!\ \cite{foscarin2024beatthisaccuratebeat}. We presented the architectural modifications required to transform the original model into a causal variant suitable for low-latency online inference, including the conversion of network layers to causal forms, strategies for reducing latency, and a new causal postprocessing method, Shifted Causal Local Max.

The results demonstrate that the original architecture can be successfully adapted for online inference while preserving much of its performance. Although the introduction of causality inevitably reduces the temporal context available to the model, when compared to existing online approaches, the adapted architecture demonstrates state-of-the-art performance on the test dataset.

The proposed Shifted Causal Local Max method proved particularly effective in addressing the limitations of strictly causal peak detection. By training the network to predict one or several frames ahead, the method effectively introduces a small lookahead window that improves peak picking without violating causal constraints and enables more reliable beat and downbeat detection compared to standard causal postprocessing methods.

While we demonstrate that in principle, our system can process audio 10--20 times faster than real time, an actual real-time implementation was outside the scope of this work, and still requires substantial engineering effort: to process individual incoming audio frames with minimal overhead, the intermediate representations of the model's convolutions and transformers need to be cached and reused, ideally in a compiled language.

To foster building on our experiments -- or implementing a real-time version -- we publish our code and model weights.\footnote{\url{https://github.com/CPJKU/beat_this_live}}

\section{Acknowledgements}
This work was supported by the European Research Council (ERC) under the EU’s Horizon 2020 research \& innovation programme, grant agreement No.\ 101019375 (Whither Music?), and the Federal State of Upper Austria (LIT AI Lab).

\bibliography{references}

\end{document}